# Quantum enhanced X-ray detection

S. Sofer[1,2], E. Strizhevsky[1,2], A. Schori[1,2], K. Tamasaku[2], and S. Shwartz[1,2*]

[1]*Physics Department and Institute of Nanotechnology, Bar-Ilan University, Ramat Gan, 52900 Israel*

[2]*RIKEN SPring-8 Center, 1-1-1 Koto, Sayo-cho, Sayo-gun, Hyogo 679-5148 Japan*

We present the first experimental demonstration of quantum-enhanced detection at x-ray wavelengths. We show that x-ray pairs that are generated by spontaneous down-conversion can be used for the generation of heralded x-ray photons and measure directly the sub-Poissonian statistics of the single photons by using photon number resolving detectors. We utilize the properties of the strong time-energy correlations of the down converted photons to demonstrate the ability to improve the visibility and the signal-to-noise ratio of an image with a small number of photons in an environment with a noise level that is higher than the signal by many orders of magnitude. In our work we demonstrate a new protocol for the measurement of quantum effects with x-rays using advantages such as background free measurements that the x-ray regime offers for experiments aiming at testing fundamental concepts in quantum optics.

*Sharon.shwartz@biu.ac.il

One of the most important results of the development of quantum optics is the ability to use quantum states of light to improve the quality of measurements with respect to conventional classical coherent or incoherent illumination [1-10]. Examples for methods that are based on quantum states of light are quantum imaging [2-5] and quantum metrology [8-10].

The extension of concepts of quantum optics to the x-ray range of the electromagnetic spectrum can lead to a new paradigm that can be utilized to test those concepts by using the advantages of the x-ray range. Examples for those advantages include the availability of detectors with true capabilities to resolve the number of photons, the nominally zero background noise, and quantum efficiency that is practically unity in a very broad spectral range. In addition, the short wavelengths of x-rays allow the access to atomic scale phenomena and can open the possibility to test concepts of quantum optics in the microscopic world. X-ray measurements could benefit from concepts of quantum optics especially when low radiation dose measurements are required or when the reflection of the sample is weaker than the scattering from the surrounding environment. Another interesting potential direction is the possibility to couple single x-ray photons with Mössbauer nuclei as proposed in many recent publications [11-22]. We note that several quantum effects with x-rays have been proposed and analyzed [11-24]. The ability to control single γ-photons emitted from Mossbauer nuclei has been observed recently [11].

However, the extension of quantum optics to the x-ray regime requires overcoming many challenges. One of the main challenges is the challenge to generate entangled photons at high flux at those wavelengths. Similar to the optical regime, one of the potential sources is the nonlinear effect of parametric down-conversion (PDC), in which pump photons interact with vacuum fluctuations in a nonlinear crystal to generate entangled photon pairs, denoted as signal and idler photons. However, the nonlinearity in the x-ray regime is significantly lower than the nonlinearity in the optical regime thus the realization of PDC with optical radiation is more available and widespread.

X-ray PDC has been demonstrated by several authors [25-30] and the application of the effect as a source for ghost imaging has been demonstrated recently [30]. However, in all previous publications, the photon statistics have not been measured. Essentially, to date, there are no experimental evidence that photons, which are generated by x-ray PDC, exhibit statistics of quantum states of radiation. Likewise, observations of the quantum-enhanced measurement sensitivity have never been reported at x-ray wavelengths.

Here we show for the first time that x-ray pairs that are generated by x-ray PDC can be used for the generation of heralded photons with perfect sub-Poissonian statistics. We demonstrate the improvement of the visibility and the signal-to-noise ratio (SNR) by using the strong time-energy correlations of photon pairs that are generated by x-ray PDC. Our protocol is similar to the protocol of quantum illumination where entanglement between two photons is utilized for the detection of objects in a very noisy environment [6]. In the quantum illumination protocol a signal photon, which probes the object, is entangled with an ancilla photon, which is retained by the user. The detection of the object is done by correlating the signal photons with the ancilla photons. Since the ancilla and the signal photons were born entangled, the strong correlations between them can be used to improve the signal-to-noise-ratio (SNR) even when the entanglement at the detectors is lost due to the noisy environment [6].

While we use the strong time-energy correlation between the ancilla and signal photons the in our detection scheme, the time and energy resolutions of our setup are insufficient to prove that the generated photon pairs are entangled since the bandwidth of the generated pairs is on the order of keV and the corresponding biphoton correlation time is on the order of a few attoseconds. However, the theory for x-ray PDC predicts that the photon pairs are time-energy entangled and we essentially observe the same results of SNR enhancement as in the protocol of quantum illumination even without proving entanglement.

**We observe a clear enhancement of the SNR relative to classical measurement methods. We show that the improvement in the SNR occurs only when we observe true sub-Poissonian statistics of the measured photons. This is a clear evidence for the quantum nature of the photon pairs, which is the reason for the enhancement of the SNR in our experiment.**

We conducted the experiment described below at the RIKEN SR physics beamline (BL19LXU) of SPring-8 [31]. The schematic of the experimental system is shown in Fig. 1. We use a pump beam at 22.3 keV to generate the photon pairs via x-ray PDC in a nonlinear diamond crystal. The photon pairs generated by PDC conserve energy, such that $\omega_p = \omega_s + \omega_i$ where $\omega_p$, $\omega_s$, and $\omega_i$ are the angular frequencies of the pump, signal, and idler, respectively. The momentum conservation condition (phase matching condition) has to be achieved, and can be written as $\vec{k}_p + \vec{G} = \vec{k}_s + \vec{k}_i$ where $\vec{k}_p$, $\vec{k}_s$, and $\vec{k}_i$ are the wave vectors of the pump, signal, and idler, respectively, and $\vec{G}$ is the reciprocal lattice vector of the nonlinear crystal. One of the emerging photons is collected by a detector that we denote as the ancilla detector (which collects the idler photons) and we use it as a trigger for the second detector. The second photon, which we denote as the signal photon is collected by a second detector. The detectors we use are silicon drift detectors that provide a signal, which is proportional to the photon energy for every of the detected photons, thus we can resolve the number of the detected photons as well as their photons energies. There unique capabilities allow us to measure background free single photons with high probability. The signal photons are registered by our data collection system only when the ancilla photon is detected. Hence the signal photons are heralded and have the prosperities of true single photons.

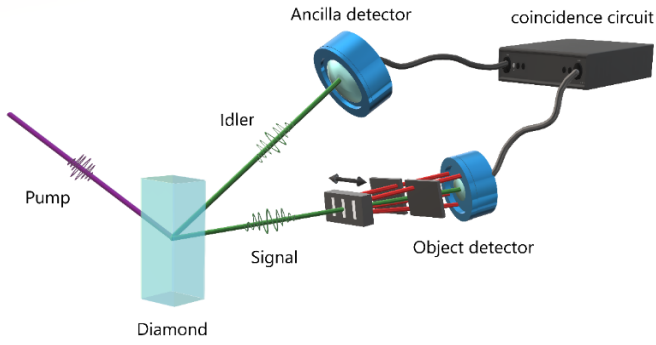

**Figure 1**: Scheme of the experimental setup. The purple beam is the pump, the green beams are the signal and idler, respectiveley, and the red beams represent the noise radiation. The object is made ot three slits. The detectors are silicon drift detetcors.

Our protocol for the measurement of the heralded photons includes two steps. In the first step we collect the photon pairs including their photon energies. In the second step we scan the data and register only photons that satisfy energy conservation by testing that the sum of the two detected photons is equal to the photon energy of the pump photon (within the resolution of our system, which is about 500 eV).

Since the abilities to generate and detect single photons are essential for most quantum schemes, we start by showing that the heralding procedure we perform indeed leads to the observation of exactly one photon at every detection event. We note that this is in contrast to measurements with low dose classical radiation where the average number of photons can be on the order of one photon or less, but the variance is large. The ability to determine exactly the number of photons and not just the average of the number of photons is a key difference between classical and quantum illuminations. Figure 2 shows the measured probability distributions of the detected events at the object and the ancilla detectors where the time window for a measurement cycle is 1 μs. We plot the probability distribution for two cases: in 2A we do not apply any post-selection energy filters, thus count mostly noise (classical radiation). In 2B we apply the energy conservation by using the energy resolving capabilities of the detectors and observe only PDC events (quantum radiation). **It is clear from Fig. 2B that as we expect, the photons that are generated by x-ray PDC are generated always in pairs. The implication is that once the ancilla photon is detected, there is exactly one correlated photon (that according to theory was born entangled with the ancilla photon) in the system.** Hence, the photons at the object detector are true single photons, which obey sub-Poissonian statistics.

A common criterion to determine whether a source is classical or quantum is the degree of correlation between the two beams [2, 4, 32-36], which is defined as

$$\sigma = \frac{\langle \delta^2(N_s - N_i) \rangle}{\langle N_i + N_s \rangle} \quad (3),$$

where $N_s$ and $N_i$ are the number of signal and idler (ancilla) photons, respectively, and delta is the variance. The degree of correlation is larger than unity for classical sources and smaller than unity for quantum sources with optical radiation. In the x-ray regime, however, due to the photon-number-resolving capabilities of detectors it is possible to decrease the degree of correlation below unity even for classical sources. This is done by post processing where we register only events where at least one photon is collected by each of the detectors in the time window of a single coincidence measurement. This method filters out events where there are no detected photons at one of the detectors. Since the coincidence time window is short, the probability to measure more than one photon at each of the detectors within this window is very low. Thus, in most of the registered events there is exactly one photon at each of the detectors and the average degree of correlation can be smaller than unity. The average degree of correlation obtained for the classical measurement in our experiment is $\sigma \sim 0.25$ while for the quantum measurement, the degree of correlation vanishes ($\sigma = 0$). We note that the degree of correlation is zero for the quantum radiation, since the quantum efficiency of x-ray detectors are nominally unity and since the dark count rate is negligible [2,4].

This is a remarkable result that demonstrates the advantages of the measurements of quantum effects with x-rays. Due to the commercially available x-ray detectors with a quantum efficiency near unity, the ability to measure the sub-Poissonian statistics of quantum x-ray radiation can be followed by demonstration of various experiments in quantum optics with nominally zero background. This is in contrast to most of the demonstrations of quantum effects in the optical regime, where the ability to resolve the number of photons is limited and the measurements are accompanied with noise.

After concluding that the pairs generated by PDC exhibit quantum properties, we demonstrate the ability to enhance the SNR with quantum radiation. Since we use a single-pixel detector for the object detector we reconstruct the image by scanning the object with respect to a slit, which we mount at the center of the detector. The object detector is exposed to a high level of noise originated from x-ray fluorescence and Compton scattering. The total noise in the energy range of the PDC process is about 4 orders of magnitude stronger than the PDC signal. We note that it is very hard to filter out this noise with classical detection since it has a significant contribution from the object itself.

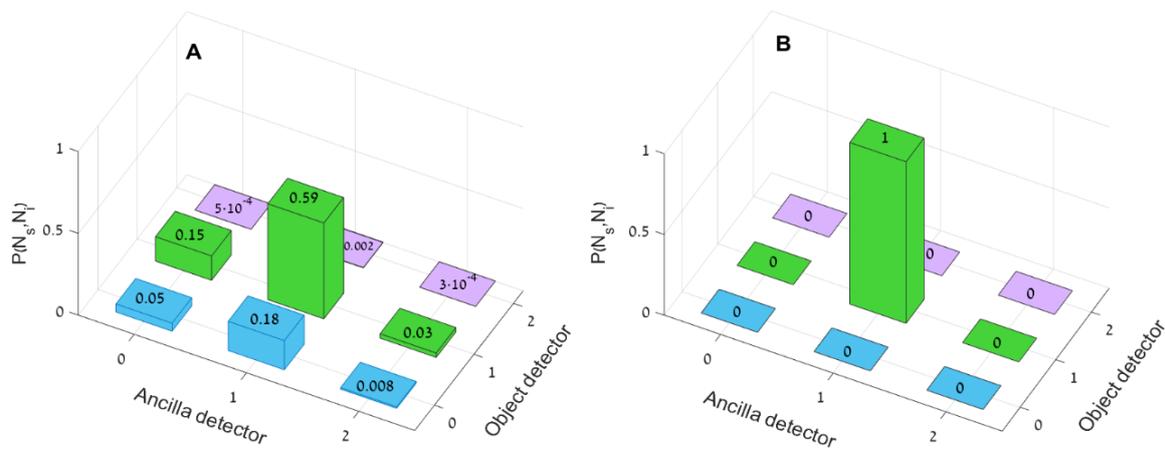

Figure 2: Probability distribution of the number of photons that are detected by the ancilla (signal) and object (idler) detectors by using coincidence measurements with a time window of 1 μs for (A) classical source and (B) PDC quantum source. The x-axis represents the number of the detected ancilla photons, and the y-axis represent the number of the detected signal photons, both in a time window of 1 μs.

In Fig. 3A we show the images obtained by using the quantum detection procedure. For the comparison, we show the images that are obtained by classical measurements with a comparable average number of photons (3B, 3C) (~100 photons per position). The image in Fig. 3B is obtained by illuminating the object with classical radiation and measuring the intensity only at the object detector. We obtain the image in Fig. 3C by using coincidence measurements between the ancilla and object detectors with classical radiation (only ~10% of the signal is originated from PDC). The advantages of the quantum scheme over the classical schemes as indicated from the comparison of Fig. 3A with Figs. 3B and 3C are prominent.

The figure of merit for the quality of the images is their visibility, which is defined as

$$v = \frac{\langle I_{max}\rangle - \langle I_{min}\rangle}{\langle I_{max}\rangle + \langle I_{min}\rangle} \quad (1),$$

where $\langle I_{max}\rangle$ and $\langle I_{min}\rangle$ are the ensemble average of the intensities above and below the chosen threshold, respectively. The visibility for the classical methods shown in Fig. 3(B, and C) are $v_B = 0.225 \pm 0.001$ and $v_C = 0.432 \pm 0.004$, respectively, while the visibility for the quantum-enhanced photo-detection scheme is $v_A = 0.998 \pm 0.002$. Clearly, the visibility obtained with the quantum-enhanced photo-detection scheme is significantly higher than the classical methods and approaches unity.

Next, we compare the SNR, which is defined as

$$SNR = \frac{\langle I_{max}\rangle}{\langle I_{min}\rangle} \quad (2).$$

We find that the SNR for the classical radiation schemes shown in Figs. 3B and 3C is 1.5 and 2.5, respectively, and that the SNR for the quantum detection scheme is on the order of $10^3$. In other word, the SNR of the quantum enhanced photo detection scheme is about three orders of magnitude higher than the SNR of classical detection schemes. This is mainly because the quantum detection scheme is very efficient in eliminating random detections events, which are the dominant sources for the background noise in our experiment.

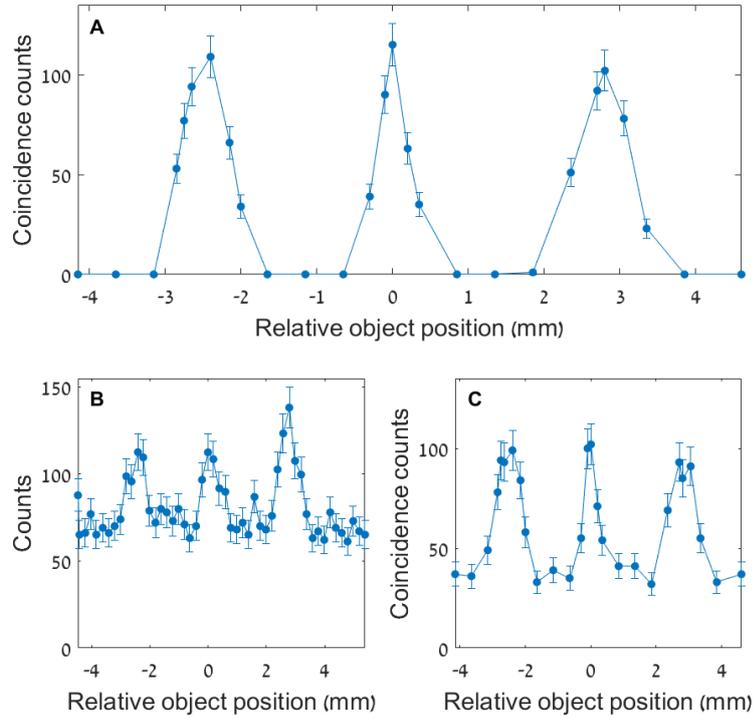

Figure 3: Reconstruction of the image of the triple slit object by (A) quantum radiation. (B) Classical radiation. (C) Classical coincidence counting. The average number of counts is comparable in all the panels. In each of the panels, the horizontal axis represents the relative position of the object and the vertical axis represents the number of events that are detected by the detection system. The error bars are estimated by assuming a Poisson distribution. The solid lines are guides for the eye.

Of importance, we note that although we measured a degree of correlation below unity also for the classical radiation with post-selection, the contrast of the reconstructed image is improved only when the degree of correlation approaches zero, which occurs only for the case of the true heralded photons. The main difference between the classical and quantum radiation is that in the classical case the detectors collect photons that are not correlated with the transmission of object since they are originated from photons that do not interact with the object or from fluorescence from the object itself, which is actually stronger at areas where the transmission is smaller. Conversely, the very efficient scheme of quantum detection registers only photons that are correlated with the transmission of the object.

In conclusion, we have demonstrated the generation and application of heralded x-ray photons generated by x-ray PDC. By using photon number resolving detectors with zero dark counts we have shown that the heralded photons obey the sub-Poissonian statistics of ideal single photons with zero background. We have demonstrated the ability to utilize the strong time-energy correlations of photons pairs for quantum enhanced photo detection. The procedure we have presented possesses great potential for improving the performances of x-ray measurements. We anticipate that this work will open the way for more quantum enhanced x-ray regime detection schemes including the area of diffraction and spectroscopy. These methods can be extremely useful for the measurement of weak signals.

Finally, we note that new x-ray sources such as high repetition rate x-ray free electron lasers can provide much higher flux than we used in the present experiment [37]. The use of seeded x-ray free electron lasers will open even more possibilities since those source exhibit correlation properties with are equivalent to optical lasers [38]. Hence the observation of quantum effects with those sources with much higher output yields is very likely.

**Acknowledgements:**

We acknowledge the SPring8 Facility for provision of synchrotron radiation facilities. This work was supported by the Israel Science Foundation (ISF) (IL), Grant No. 201/17

## Appendix I: experimental details

We provide further information on the experimental setup.

The input power is about $5\times10^{13}$ photons/sec. The input beam is monochromatic at 22.3 keV and is polarized in the scattering plane. The dimensions of the beam at the input are 0.2 mm (vertical) × 0.5 mm (horizontal). The nonlinear crystal is a diamond crystal with dimensions of 4 mm × 4 mm × 0.8 mm. We use the <660> reciprocal lattice vector in Laue geometry to achieve phase matching. The Bragg angle is 41.5º. The deviation from the Bragg angle for the phase matching condition is 10 mdeg. The angular separation between the ancilla and object detectors is ~2.06º. We use a helium duct between the crystal and the detectors in order to reduce air scattering and air absorption of the generated PDC pairs. In order to achieve spatial resolution, we mount a 0.5 mm slit before the object detector and scan by shifting the object with respect to the detector. This procedure ensures that the PDC spectra are constant at all object positions.

Next, we discuss the coincidence electronics. Both the ancilla and object detectors provide two output pulses upon the detection of a photon. One of the output pulses from the detectors is a logical pulse with the duration of 1 μs. Both logical pulses from the detectors are used as inputs for an AND gate, which provides 1 μs long trigger pulses. The inverse peaking time of the detectors, which is about 100 ns. The second type of pulses the detectors provide is analog, where the heights of the pulses are proportional to the energy of the detected photons. The calibration of the photon energy is done by measuring the voltage that corresponds to the input photon energy at 22.3 keV. This scheme allows us to record events where at least one photon arrives at each of the detectors within the time window we set. The energies of the detected photons are also measured in this scheme. In order to post-select only the events that are originated from PDC, we register only events when the sum of the photon energies of the signal and idler photons of each of each of the pairs is equal to the photon energy of the input photon and that arrive with a time difference of no more than 250 ns between the two detectors. We also choose a bandwidth of 2 keV around the degeneracy energy for the two detectors. To calculate the visibility and SNR we consider photon counts that are 30% below than the maxima as the maxima and photon counts that are 30% above the minimal count as the minima.

## Appendix II: Verification of the PDC source

We verify that we indeed measure PDC by measuring the spectra without the object for several detector angles [28-30]. Since there is a one-to-one relation between the photon energies and the angles of propagation of the signal and idler photons, which are determined by the phase matching equation, we expect to measure different spectra for different detector angles. Figures A1, A2 show the measured spectra for the degenerate case while Figs. B1-D2 show the spectra for different offsets from the degenerate phase matching solution. The shift of the spectra we measured agrees with our simulations and with the phase matching calculations. The measured coincidence rate is ~100 coincidence counts/hour, which is about five orders of magnitude larger than the expected coincidence rate from accidental noise photons.

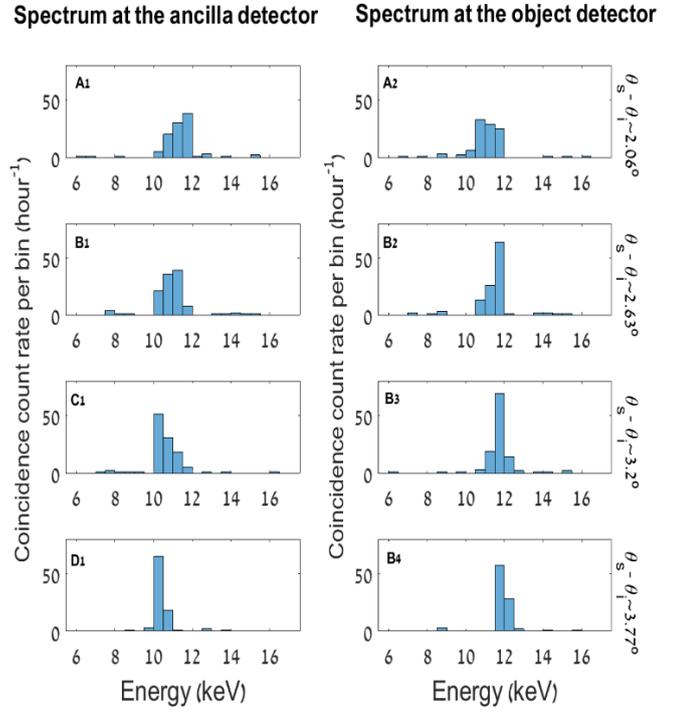

**Figure 4:** PDC spectra for the ancilla (left plot) and object detector (right plot) for angular separation between the detectors of (A1, A2) ~2.06º, degenerate case, (B1, B2) ~2.63 º, (C1, C2) ~3.2 º, and (D1, D2) ~3.77 º.